\documentclass[prd,tightenlines,nopacs,nofootinbib,eqsecnum,amsfonts,amssymb,amsmath]{revtex4}
\pdfoutput=1
\usepackage{amssymb,amsmath,amsfonts,amsthm,graphicx,bm}

\def\be{\begin{equation}}
\def\ee{\end{equation}}
\def\bea{\begin{eqnarray}}
\def\eea{\end{eqnarray}}

\def\e{{\bm e}}
\def\sp{\sigma_+}
\def\udot{\dot{u}}
\def\ex{\alpha_\|}
\def\ey{\alpha_\perp}
\def\ez{e_3{}^3}
\def\R{{}^3\!R}
\def\S{{}^3\!S_+}
\def\eyinv{A_\perp}
\def\apey{(a+e_2{}^2)}
\def\apey{(a+\ey)}

\def\y{\vartheta}

\def\ra{\sqrt{A}}

\begin{document}

\title{Spherically symmetric cosmological spacetimes with dust and radiation\\ -- numerical implementation -- }

\author{Woei Chet Lim$^{1,2}$, Marco Regis$^{3}$ and Chris Clarkson$^{4}$}
\affiliation{$^{1}$ Department of Mathematics, University of Waikato, Private~Bag~3105, Hamilton 3240, New Zealand\\
$^{2}$ Max-Planck-Institute for Gravitational Physics, Am~M\"{u}hlenberg~1, D-14476 Potsdam, Germany\\
$^{3}$ Dipartimento di Fisica, Universit\`a  di Torino and INFN, Torino, Italy\\
$^{4}$ Astrophysics, Cosmology and Gravity Centre, and Department of Mathematics and Applied Mathematics, University of Cape Town, Rondebosch 7701, Cape Town, South Africa}
\email{wclim@waikato.ac.nz, regis@to.infn.it, chris.clarkson@gmail.com}

\begin{abstract}
We present new numerical cosmological solutions of the Einstein Field Equations. The spacetime is spherically symmetric with a source of dust and radiation approximated as a perfect fluid. The dust and radiation are necessarily non-comoving due to the inhomogeneity of the spacetime. Such a model can be used to investigate non-linear general relativistic effects present during decoupling or big-bang nucleosynthesis, as well as for investigating void models of dark energy with isocurvature degrees of freedom.  We describe the full evolution of the spacetime as well as the redshift and luminosity distance for a central observer.
After demonstrating accuracy of the code, we consider a few example models, and demonstrate the sensitivity of the late time model to  the degree of inhomogeneity of the initial radiation contrast.
%PACS numbers: 98.80.Jk, 04.20.-q, 04:20.Jb, 04.25.D-
\end{abstract} 
\maketitle

\section{Introduction}

We have developed a code to fully solve for the dynamics of a general relativistic spacetime which is inhomogeneous but spherically symmetric, with a source given by matter and radiation.\footnote{COSMO-DRESS
(COSMOlogical Dust and Radiation Evolution in a Spherically Symmetric spacetime), is available on request.} Exact solutions of the Einstein Field Equations exist for such a geometry if the matter is purely dust (Lema\^\i tre-Tolman-Bondi models), or pure radiation (so called Lema\^\i tre models)~\cite{Bolejko:2010zz}, but in the case where both sources are present an exact solution is not possible in general (see also \cite{Marra:2011zp,Marra:2012pj} for the set of equations in the case of $n$ perfect fluids). Because the dust is geodesic and the radiation not (except in the homogeneous sub-case) the dust and radiation are non-comoving, making the total effective fluid anisotropic. 

There are a variety of circumstances where a model such as this is important to investigate, for example:

\begin{description}

\item[Relativistic effects during decoupling] The non-linear evolution of cosmological perturbations during decoupling leads to important deviations from Gaussianity in the CMB as well as the late time matter power spectrum (see e.g.~\cite{Pitrou:2008ak}). Furthermore, non-linear dynamics of the photon-baryon plasma coupled through gravity to the dark matter will lead to subtle changes to the baryon acoustic oscillations and the corresponding peak in the galaxy power spectrum. General relativistic aspects of deviations from canonical linear theory can be studied in a model setup such as this, and can be used to place limits on the accuracy of perturbation theory, although the simplification of spherical symmetry means that we could not produce a comprehensive measurement prediction. 

\item[Void models of dark energy] One of the key constraints on void models for dark energy (see~\cite{Clarkson:2012bg} for a comprehensive review) arise from combined constraints from the CMB combined with local measurements of $H_0$ (see, e.g., \cite{Moss:2010jx}). It was argued in~\cite{Clarkson:2010ej} that those constraints only apply to adiabatic voids in which the radiation  is assumed to be homogeneous while that of the matter is inhomogeneous. Furthermore, observations such as the kinetic Sunyaev-Zel'dovich effect and the baryon acoustic oscillation feature in the matter power spectrum depend strongly on the decoupling time; as this may not be homogeneous, predictions of these effects can alter dramatically depending on the initial radiation profile. To understand this we need a fully relativistic modelling of a spherically symmetric spacetime which allows for non-comoving matter and radiation at the background level. 

\item[Inhomogeneous Big-Bang Nucleosynthesis] The possibility of an inhomogeneous Big-Bang nucleosynthesis (BBN) with a varying baryon to photon and/or neutron to proton ratio on scales comparable to particle diffusion horizon at BBN has been widely studied in the literature, with the inhomogeneities often modelled assuming spherical symmetry (see, e.g., \cite{Lara:2006cd} for a review).
The solution presented here can be exploited to study the case of isocurvature inhomogeneities and to consistently couple the particle physics dynamics to the expansion of the locally inhomogeneous Universe.

\end{description}

In the first iteration of the model presented here we make two approximations.
The baryons and dark matter are described as a single dust fluid, which is a poor approximation during decoupling when the baryons are tightly coupled to the photons through Compton scattering.
The radiation is treated as a perfect fluid. In general however, in  the Einstein-Boltzmann equations for the radiation fluid, higher moments such as the anisotropic pressure cannot be neglected in an inhomogeneous spacetime.

Our current focus is however on a model in which the length-scale associated with the spherical inhomogeneity is much larger than the horizon size at matter-radiation equality. 
This allows us to disregard the dragging of baryons by the photons until decoupling.
Moreover, since such models develop significant shear only at late times and on large scales,
neglecting the higher moments should be a reasonable approximation to begin with (anisotropic pressure is sourced
by the energy density of the radiation times the shear). 

Both of these approximations will be removed in later iterations of the code.

The paper is organized as follows. In Section~\ref{sec:model} we present the system of equations describing a spherically symmetric spacetime with a source of dust and radiation. The numerical implementation is described in Section~\ref{sec:num}. In Section~\ref{sec:ex} we show concrete physical examples which are also used to test the accuracy of the code. Section~\ref{sec:conc} concludes.

\section{The model} \label{sec:model}

\subsection{Spherically Symmetric Einstein Field Equations}

We shall write the Einstein field equations using the orthonormal frame formalism~-- see~\cite{vanElst:1996dr,book:WainwrightEllis1997} which we refer to for definitions and notation.
The original derivation can be found in~\cite{Coley:2008zz,Coley:2008qd}, which we reproduce here.
The line-element has the form
\be
	ds^2 = - N^2 d\tau^2 + (\ex)^{-2} dr^2 + (\ey)^{-2} (d\vartheta^2 + \sin^2 \vartheta d\varphi^2).
\ee
$\ex$ and $\ey$ are frame coefficients and depend on $\tau$ and $r$. (We reserve $t$ for proper time.)
The Killing vector fields are given by \cite{book:Stephanietal2003}:
\be
    \partial_\varphi,\quad
    \cos \varphi \ \partial_\vartheta - \sin \varphi \cot \vartheta \ \partial_\varphi,\quad
        \sin \varphi \ \partial_\vartheta + \cos \varphi \cot \vartheta \ \partial_\varphi.
\ee
The orthonormal frame vectors are:
\be
    \e_0 = N^{-1} \partial_\tau
    ,\quad
    \e_1 = \ex \partial_r
        ,\quad
        \e_2 = \ey \partial_\vartheta
        ,\quad
        \e_3 = \ez \partial_\varphi,
\ee
where $\ez = \ey / \sin \vartheta$. $N$, $\ex$ and $\ey$ are functions of $\tau$
and $r$.

Spherical symmetry leads to the following restrictions on the orthonormal frame components of the kinematic variables~-- the shear, vorticity and acceleration:
\be
    \sigma_{\alpha\beta} = \text{diag}(-2\sp,\sp,\sp),\quad
    \omega_{\alpha\beta} =0,\quad
    \udot_\alpha =(\udot_1,0,0),
\ee
where
\be
    \udot_1 = \e_1 \ln N.
\ee
The spatial commutation functions simplify to:
\be
    a_\alpha = (a_1, a_2, 0),\quad
    n_{\alpha\beta} = \left( \begin{array}{ccc}
            0 & 0 & n_{13}  \\
            0 & 0 & 0   \\
            n_{13} & 0 & 0 \end{array} \right),
\ee
where
\be
    a_1 = \e_1 \ln\ey,\quad
    a_2 = n_{13} = - \frac12 \ey \cot \y;
\ee
and the matter components simplify to:
\be
    q_\alpha = (q_1,0,0),\quad
    \pi_{\alpha\beta} = \text{diag}(-2\pi_+,\pi_+,\pi_+).
\ee
The frame rotation $\Omega_{\alpha\beta}$ is zero. To compare with the setup in~\cite{Clarkson:2010ej}, we identify:
\begin{align}
	N &= e^\phi,
\\
	\ey &= \frac{1}{a_\perp r},
\\
	(\ex)^{-2} &= \frac{a_{||}^2}{1-kr^2},
\\
	a &= - \frac{\sqrt{1-kr^2}}{a_\perp r},
\\
	H+\sp &= \tilde{H}_\perp,
\\
	H-2\sp &= \tilde{H}_{||};
\end{align}
(see \cite{Clarkson:2010ej} for the notation).
        
Now, $n_{13}$ only appears in the equations together with
$\e_2 n_{13}$ in the form of the Gauss curvature of the spheres
\be
    {}^2\!K := 2(\e_2 - 2 n_{13}) n_{13},
\ee
which simplifies to
\be
    {}^2\!K = (\ey)^2.
\ee
Thus there is no dependence on $\vartheta$ in the equations.

The spatial curvatures also simplify to:
\be
        {}^3\!S_{\alpha\beta} = \text{diag}(-2\S,\S,\S),
\ee
with $\R$ and $\S$ given by:
\begin{align}
        \R &= 4 \e_1 a_1 - 6 a_1^2 + 2 {}^2\!K
\\
        \S &= - \tfrac13 \e_1 a_1 + \tfrac13 {}^2\!K.
\end{align}
   
We will drop the index from
$\udot_1,\ a_1$ and write them as 
$\udot,\ a$.
To summarize, so far there are ten variables
\be
    N,\ \ex,\ \ey,\ H,\ \sp,\ a,\ \rho,\ q_1,\ p,\ \pi_+.
\ee
The evolution of $N$, $p$ and $\pi_+$ are yet to be specified.
A temporal gauge choice determines $N$, and a fluid model determines $p$ and $\pi_+$.

Before the matter is specified and the gauge chosen, we now have the following underdetermined system.
\begin{align}
    \e_0 \ex &= (-H+2\sp) \ex
\\
    \e_0 \ey &= -(H+\sp) \ey
\\
    \e_0 H &= - H^2 - 2 \sp^2  + \tfrac13 (\e_1 + \udot - 2 a)\udot - \tfrac16(\rho+3p)
\\
    \e_0 \sp &= -3H \sp - \tfrac13(\e_1 + \udot + a)\udot
        - \S + \pi_+
\\
    \e_0 a &= (-H+2\sp) a - (\e_1 + \udot)(H+\sp)
\\ 
    \e_0 \rho &= -3H(\rho+p) - (\e_1+2\udot-2a)q_1 - 6\sp\pi_+
\label{B1}
\\
    \e_0 q_1 &= (-4H+2\sp)q_1 - \e_1 p -(\rho+p)\udot
            + 2(\e_1+\udot-3a)\pi_+.
\label{B2}
\end{align}
%where 
%\be
%        \e_0 = N^{-1} \partial_\tau,\quad
%        \e_1 = \ex \partial_r.
%\ee
The constraint equations are the Gauss and Codazzi constraints, and the
definition of $a$:
\begin{align}
    0 = (C_G) &= 3H^2 + \tfrac12 \R - 3 \sp^2 - \rho
\\
    0 = (C_C)_1 &= -2 \e_1(H+\sp) + 6 a \sp + q_1
\\
    0  = (\text{def $a$}) &= (\e_1 - a) \ey,
\end{align}
where the spatial curvatures are given by
\begin{align}
    \R &= 4 \e_1 a - 6 a^2 + 2 (\ey)^2
\\
    \S &= - \tfrac13 \e_1 a + \tfrac13 (\ey)^2.
\end{align}

\subsection{The matter: dust plus radiation}

We shall study models with two perfect fluids (dust and radiation) with stress energy tensors
\begin{eqnarray}
T_{ab}^\text{dust}=\rho u^\text{dust}_a u^\text{dust}_b,\\
T_{ab}^\text{rad}=\tfrac{4}{3}\mu^\text{rad} u^\text{rad}_a u^\text{rad}_b+\tfrac{1}{3}\mu^\text{rad} g_{ab}.
\end{eqnarray}
  We require that each fluid satisfy the twice-contracted Bianchi identities 
(\ref{B1})--(\ref{B2}). We choose the temporal gauge to be comoving with the dust: $\bm u^\text{dust}=\e_0$, and set the lapse to $N=e^\tau$.
We use $\rho$ to denote the dust density,  and $v$ for the radiation tilt:
\be
\bm u^\text{rad}=\gamma (\e_0+v\e_1),~~~~\gamma=(1-v^2)^{1/2}.
\ee
then the radiation density measured in the dust frame is given by
\be
\mu=\mu^\text{rad}\left(1+\tfrac{4}{3}\gamma^2 v^2\right).
\ee

After the matter is specified and the gauge chosen, we now have the following closed system:
\begin{align}
\label{evo_ex}
    \e_0 \ex &= (-H+2\sp) \ex
\\
    \e_0 \ey &= -(H+\sp) \ey
\\
    \e_0 H &= - H^2 - 2 \sp^2 - \tfrac16\rho - \tfrac13\mu
\\
    \e_0 \sp &= -3H \sp - \S  - \frac{4 \mu}{9 G_+} v^2
\label{e0_sp}
\\
    \e_0 a &= (-H+2\sp) a - \e_1 (H+\sp)
\label{e0_a}
\\ 
        \e_0 \rho &= - 3 H \rho
\\
\label{e0_mu}
    \e_0 \mu &= - \frac{4 v}{3G_+} \e_1 \mu
        - \frac{4 G_-}{3G_+^2} \mu \e_1 v
        - \frac{4}{3G_+} \mu \left[ 
            (3+v^2)H - 2 v a - 2 v^2 \sp \right]
\\ 
\label{e0_v}
    \e_0 v &= - \frac{(1-v^2)^2}{4 G_- \mu} \e_1 \mu
    - \frac{8v^3}{9 G_+ G_-} \e_1 v
        - \frac{(1-v^2)v}{G_-} \left[ - 2 \sp + \frac23 v a \right],
\end{align}
where $G_\pm = 1 \pm \frac13 v^2$.  
The constraint equations are
\begin{align}
    0  &= 3H^2 + \tfrac12 \R - 3 \sp^2 - \rho - \mu
\\
    0  &= -2 \e_1(H+\sp) + 6 a \sp + \frac{4 \mu}{3 G_+} v
\\
    0  &= (\e_1 - a) \ey.
\end{align} 

We have made two key assumptions in the above equations: that the radiation is a perfect fluid, and that the baryons and dark matter can be treated as a single dust fluid, which is only valid in the free-streaming case.

In order to relax the perfect fluid approximation for radiation mentioned in the Introduction, one should add the evolution of the quadrupole and higher multipoles (see, e.g., Section VI in \cite{Maartens:1998xg} for the 
equations). It implies modifying the evolution equation for the radiation density, which would receive a source term from the anisotropic pressure part of the radiation (the quadrupole); the evolution equation for this is sourced by the octopole, and so on.
This significantly increases the complexity of the system (and increases the number of terms which need to be regularized or kept with small numerical errors), and will be left to future iterations of the code. 

To describe baryons and dark matter as two separate, non-comoving, fluids, two new variables (a density and relative velocity) enter the system, adding two evolution equations and a few extra terms to the other equations. The appropriate terms for coupling baryons to photons come from the Boltzmann equation (again see, e.g., \cite{Maartens:1998xg}).  (At very early times when the baryons and dark matter become relativistic approximating these components as dust breaks down.)
Including these effects will also be left to future versions of the code.

%Relaxing the perfect fluid approximation for radiation is the subject of our next iteration of the code, while we leave the latter improvement to the future.

\subsection{Distances and redshifts for a central observer.}

In this section we develop the luminosity distance and redshift relation for a dust observer at the centre to an object on the dust congruence $u_a^\text{dust}$.

Consider the observer at the centre receiving an incoming photon at the present time $t=t_0$.
Let $x_\mathrm{photon}(t)$ denote the location of the incoming photon at time $t$.
It is governed by (from $ds^2=0$)
\be
	\frac{d}{dt} x_\text{photon}(t) = - N \ex |_{(t,x_\text{photon}(t))},
\ee
with initial condition $x_\text{photon}(t_0)=0$, and $\frac{d}{dt}$ denotes the derivation along the null geodesic.

Let $k^a$ be the 4-vector of an incoming radial null geodesic ($k_ak^a=0$):
\begin{align}
        k^a = \epsilon N^{-1} \frac{\partial}{\partial t} - \epsilon \ex \frac{\partial}{\partial r} 
	 = (\epsilon,-\epsilon,0,0) \quad \text{in the orthonormal frame,}
\end{align}
which satisfies the null geodesic equation
\be
	k_{a;b} k^b = 0.
\ee
The components $k_{0;b} k^b = 0$ and $k_{1;b} k^b = 0$ both give the same equation for $\epsilon$,
the photon energy:
\be
\label{epsilon_null_geo}
        \left( N^{-1} \frac{\partial}{\partial t} - \ex \frac{\partial}{\partial r} \right) \ln \epsilon =
	N^{-1} \frac{\partial}{\partial t} \ln \ex + \ex \frac{\partial}{\partial r} \ln N.
\ee
Assuming $N=N(t)$ and using (\ref{evo_ex}), we get
%\be%
%	\left( N^{-1} \frac{\partial}{\partial t} - \ex \frac{\partial}{\partial r} \right) \ln \epsilon =
%	-  H_{\|} 
%\ee
\be
        \left( N^{-1} \frac{\partial}{\partial t} - \ex \frac{\partial}{\partial r} \right) \ln \epsilon =
        -  (H-2\sp).
\ee
For a single ray of light, we consider $ \epsilon(t,x_\text{photon}(t))$, abbreviated to $\epsilon(t)$. Then
\be
\label{evo_epsilon}
	\frac{d}{dt} \ln \epsilon(t) = - N(t)  (H-2\sp) (t,x_\text{photon}(t)).
\ee
The redshift of the incoming light,  $z$, is defined in terms of the energy of the photon by
\be
	1+z(t) = \frac{\epsilon(t)}{\epsilon(t_0)}.
\ee
The actual value of $\epsilon(t_0)$ does not matter here, so we  set it to 1 in the numerical computation.

The angular diameter distance is determined by considering the divergence of the null vector field: $\theta=\nabla_a k^a$. The expansion $\theta$ works out to be (after using (\ref{epsilon_null_geo}) to simplify):
\be
	\frac{\theta}{\epsilon} = - \left( N^{-1} \frac{\partial}{\partial t} - \ex \frac{\partial}{\partial r} \right) \ln \ey.
\ee
Let $A(t)$ denote the beam cross-section area. Then
\be
	N^{-1} \frac{d}{dt} \ln \ra = \frac{\theta(t,x_\mathrm{photon}(t))}{\epsilon(t,x_\mathrm{photon}(t))},
\ee
which implies that
\be
\label{ra_ey}
	\ra \propto \frac{1}{\ey};
\ee
$\ra$ is proportional to the angular-diameter distance $d_A$. The angular-diameter distance is then  $d_A = (1/\ey)_\text{in}$. The luminosity distance is given by $d_L = (1+z)^2 d_A$.

\section{Equations for numerical simulations} \label{sec:num}

To convert the system into a PDE system suitable for numerics there are a number of issues to consider.
% (see the Appendix for detailed information). 
%Evaluating these terms close to the origin is not as bad, but one should be careful.
First we use different variables and use the constraints to eliminate spatial derivative terms in the evolution 
equations (\ref{e0_sp}) and (\ref{e0_a})~\cite{Alcubierre:2004gn}.
Ideally one should find a change of variables which regularizes all the terms. However, it is not known whether it exists for the system under investigation, and
we failed to regularize some terms (specifically, $v/\eyinv$) in the evolution equations of the radiation fluid variables. On the other hand, the resulting equations (see below) are still more suitable for numerical evolution than 
the original equations, and we checked that numerical errors remain under control (as shown later with a few examples), so we will use them.
A coordinate singularity at origin requires us to avoid evaluating singular terms in the equations, so, as we discuss below, we will stagger the grid one-half grid width away from the origin 
(also mentioned in~\cite{Alcubierre:2004gn}).

The change of variables is the following:
\be
	\eyinv = \frac{1}{\ey},\quad
	H_{\|} = H-2\sp,\quad
	H_{\perp} = H+\sp,\quad
	\apey = a+\ey.
\ee
We refrain from assigning a new letter for the variable $\apey$.

The evolution equations (with lapse $N=e^\tau$) in $(\tau,r)$ coordinates become:
\begin{align}
\label{3.2}
        \partial_\tau \ex &= -N H_{\|} \ex
\\
        \partial_\tau \eyinv &= N H_{\perp} \eyinv
\\
        \partial_\tau H_{\|} &= -N \bigg[ H_{\|}{}^2 - H_{\perp}{}^2 + \apey^2 - 2 \frac{\apey}{\eyinv}
                        + \frac12 \rho + \frac23 \mu + \frac89 \frac{\mu}{G_+} v^2 \bigg]
\\
        \partial_\tau H_{\perp} &= -N \bigg[ \frac32 H_{\perp}{}^2 - \frac12\left( \apey^2 - 2 \frac{\apey}{\eyinv}\right)
                                        + \frac16 \mu + \frac49 \frac{\mu}{G_+} v^2 \bigg]
\\ 
        \partial_\tau \apey &= -N \left[ H_{\perp} \apey +  \frac23 \frac{\mu}{G_+} v \right]
\\
        \partial_\tau \rho &= -N (H_{\|} + 2H_{\perp})\rho
\\ 
        \partial_\tau \mu &= N \bigg[ -\frac43 \frac{v}{G_+} \ex \partial_r \mu
                                -\frac43 \frac{G_-}{G_+^2} \mu \ex \partial_r v
\notag\\
                                &\qquad -\frac43 \frac{\mu}{G_+} \left( G_+ (H_{\|} + 2H_{\perp}) - 2v\apey
                        + 2 \frac{v}{\eyinv} + \frac23 v^2(H_{\|}-H_{\perp}) \right) \bigg]
\\
        \partial_\tau v &= N \bigg[ - \frac{(1-v^2)^2}{4G_-} \ex \frac{\partial_r \mu}{\mu}
                                - \frac89 \frac{v^3}{G_+ G_-} \ex  \partial_r v
\notag\\
                        &\qquad -\frac{2(1-v^2)v}{3G_-} \left( H_{\|}-H_{\perp} + v\apey - \frac{v}{\eyinv} \right) \bigg].
\end{align}
The constraints are
\begin{align}
        0 = (C_G) &= H_{\perp}{}^2 + 2 H_{\|} H_{\perp}
                + 2 \ex \partial_r \apey + 4 \frac{\apey}{\eyinv}
                -3 \apey^2 - \rho - \mu
\\     
        0 = (C_C)_1 &= -2 \ex \partial_r H_{\perp} + 2 \apey (H_{\perp}-H_{\|})
                        - 2 \frac{(H_{\perp}-H_{\|})}{\eyinv} + \frac43 \frac{\mu}{G_+} v
\\     
\label{3.12}
        0 = (\text{def $a$}) &= \apey \eyinv + \ex \partial_r \eyinv - 1.
\end{align}

\subsection{Initial condition specification}\label{sec:IC}

    \begin{figure}
      {
        \includegraphics[width=\linewidth]{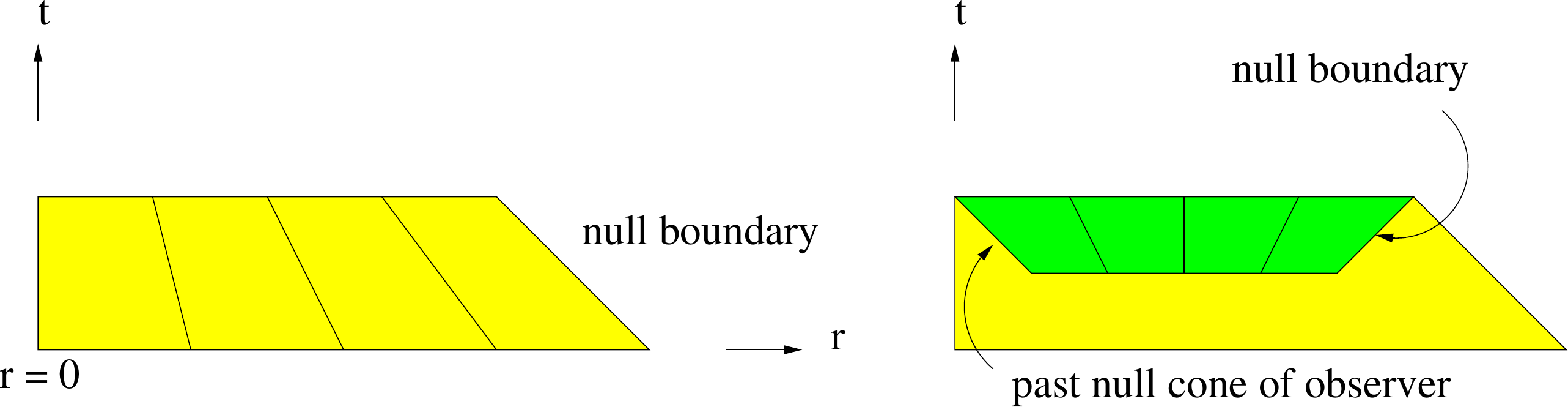}
        \caption{The spacetime simulated in the first and second stages of numerical simulation. See text in Sections~\ref{sec:IC} and~\ref{sec:evo}.}
        \label{fig:spacetime}
      }
    \end{figure}

We specify the initial condition at early times (before last scattering), and evolve the spacetime to the present day. This is the first stage of simulation.
For the sake of simplicity, we focus on simulating models with homogeneous bang time (the analysis of the impact of an inhomogeneous bang time is left for a future work).  
We thus need to specify initial condition at early times rather than at the present day in order to fulfill this requirement. 
In the second stage of the simulation, we evolve backward from present day to last scattering to obtain the redshift and luminosity distance from the centre.
%The reason for a two-stage simulation is that the null cone of the observer at present day can be identified in an evolution from the present day, but not in an evolution towards the present day.
See Fig.~\ref{fig:spacetime}.

We now describe our specification of the initial condition. 
From the evolution and constraint equations above, we see that there are eight variables and three constraints. One way to specify the initial condition is to specify five of the variables and use the constraint to compute the 
other three. For example, one can specify the four ``physical" quantities
\[
	H_{\|},\ H_{\perp}, \ \rho,\ \text{and} \ \mu,
\]
set $\ex$ (which is equivalent to a gauge choice), and then use the constraints to compute $\eyinv$,\ $\apey$, and $v$.

\subsection{Numerical Evolution and avoidance of boundary conditions}\label{sec:evo}

In this section we describe the numerical evolution in the first and second stages (see Fig.~\ref{fig:spacetime}). In the first stage we compute the spacetime region we are interested in. For this we have the following boundary 
considerations:

\begin{description}
\item[Inner boundary] We require the solution to be smooth. This implies that the variables must be either odd or even functions of $r$.
Specifically, $\eyinv$, $\apey$ and $v$ are odd, while $\ex$, $H_{\|}$, $H_{\perp}$, $\rho$ and $\mu$ are even.
This takes care of the inner boundary condition at $r=0$. 
To avoid evaluating singular terms in the evolution equations we stagger the grid from the origin; i.e. we move the grid one-half grid point from the origin.

\item[Outer boundary] We can avoid specifying outer boundary conditions by transforming to null coordinates, using the `zooming' technique in~\cite{art:Limetal2009}. To implement this, new reference coordinates $(T,X)$ are used, with $\tau=T$, and $r=r(T,X)$. 

\end{description}

The coordinate $r$ is given by the evolution equation
\be
	\partial_T r = \partial_T r_\text{in} + (\partial_T r_\text{out} - \partial_T r_\text{in}) \frac{X-X_\text{in}}{X_\text{out}-X_\text{in}},
\label{dtr}
\ee
where
\be
\label{dtr_first}
        \partial_T r_\text{in} = 0,\quad
        \partial_T r_\text{out} = - N (\ex)_\text{out},
\ee
and we choose $r=X$ initially.
By the Chain Rule, the change of coordinates from $(\tau,r)$ to $(T,X)$ changes the partial derivatives as follows:
\be
	\partial_\tau = \partial_T -  \frac{\partial_T r}{\partial_X r} \partial_X,\quad
	\partial_r = (\partial_X r)^{-1} \partial_X.
\ee
The evolution equations (with lapse $N=e^T$) in $(T,X)$ coordinates are then given by:
\begin{align}
\label{numevo_start}
	\partial_T \ex &= -N H_{\|} \ex +  \frac{\partial_T r}{\partial_X r} \partial_X \ex
\\
	\partial_T \eyinv &= N H_{\perp} \eyinv +  \frac{\partial_T r}{\partial_X r} \partial_X \eyinv
\\
	\partial_T H_{\|} &= -N \bigg[ H_{\|}{}^2 - H_{\perp}{}^2 + \apey^2 - 2 \frac{\apey}{\eyinv}
			+ \frac12 \rho + \frac23 \mu + \frac89 \frac{\mu}{G_+} v^2 \bigg] +   \frac{\partial_T r}{\partial_X r} \partial_X H_{\|}
\\
	\partial_T H_{\perp} &= -N \bigg[ \frac32 H_{\perp}{}^2 - \frac12\left( \apey^2 - 2 \frac{\apey}{\eyinv}\right)
					+ \frac16 \mu + \frac49 \frac{\mu}{G_+} v^2 \bigg] +   \frac{\partial_T r}{\partial_X r} \partial_X H_{\perp} 
\\
	\partial_T \apey &= -N \left[ H_{\perp} \apey +  \frac23 \frac{\mu}{G_+} v \right]  +   \frac{\partial_T r}{\partial_X r} \partial_X \apey
\\
	\partial_T \rho &= -N (H_{\|} + 2H_{\perp})\rho +   \frac{\partial_T r}{\partial_X r} \partial_X \rho
\\
	\partial_T \mu &= N \bigg[ -\frac43 \frac{v}{G_+} \frac{\ex}{\partial_X r} \partial_X \mu
				-\frac43 \frac{G_-}{G_+^2} \mu \frac{\ex}{\partial_X r} \partial_X v
\notag\\
				&\qquad -\frac43 \frac{\mu}{G_+} \left( G_+ (H_{\|} + 2H_{\perp}) - 2v\apey
			+ 2 \frac{v}{\eyinv} + \frac23 v^2(H_{\|}-H_{\perp}) \right) \bigg] +   \frac{\partial_T r}{\partial_X r} \partial_X \mu
\\
	\partial_T v &= N \bigg[ - \frac{(1-v^2)^2}{4G_-} \frac{\ex}{\partial_X r} \frac{\partial_X \mu}{\mu}
				- \frac89 \frac{v^3}{G_+ G_-} \frac{\ex}{\partial_X r}  \partial_X v
\notag\\
			&\qquad	-\frac{2(1-v^2)v}{3G_-} \left( H_{\|}-H_{\perp} + v\apey - \frac{v}{\eyinv} \right) \bigg] 
			+  \frac{\partial_T r}{\partial_X r}   \partial_X v.
\label{numevo_end}
\end{align}
The constraints are
\begin{align}
\label{numcon_start}
	0 = (C_G) &= H_{\perp}{}^2 + 2 H_{\|} H_{\perp} 
		+ 2 \frac{\ex}{\partial_X r} \partial_X \apey + 4 \frac{\apey}{\eyinv}
		-3 \apey^2 - \rho - \mu
\\
	0 = (C_C)_1 &= -2 \frac{\ex}{\partial_X r} \partial_X H_{\perp} + 2 \apey (H_{\perp}-H_{\|})
			- 2 \frac{(H_{\perp}-H_{\|})}{\eyinv} + \frac43 \frac{\mu}{G_+} v
\\
	0 = (\text{def $a$}) &= \apey \eyinv + \frac{\ex}{\partial_X r} \partial_X \eyinv - 1.
\label{numcon_end}
\end{align}

In the second stage we compute the distance redshift relation for an observer located at $t_\text{today}$ and $r=0$. This requires  extrapolating (with 4th order accuracy) the final data from the first stage to $r=0$, including the fractions $\frac{\apey}{\eyinv}$ and $\frac{v}{\eyinv}$.
%We require that both inner and outer boundaries be null boundaries (see Fig.~\ref{fig:spacetime}).
On the past lightcone of the central observer, the coordinate $r$ is given by the evolution equation (\ref{dtr}), where
\be
\label{dtr_second}      
	\partial_T r_\text{in} = - N (\ex)_\text{in},\quad
	\partial_T r_\text{out} = N (\ex)_\text{out}.
\ee
We also require one additional evolution equation for the redshift $z$:
\be
\label{dt_epsilon}
	\partial_T \ln (1+z) = - N H_\text{$\|$ in},
\ee
%with redshift $z$ given by $z=\epsilon-1$.
The angular-diameter distance $d_A$ is directly proportional to $(\eyinv)_\text{in}$, so we simply set $d_A = (\eyinv)_\text{in}$.

\subsubsection{Implementation}

We use the classical 4th order Runge-Kutta method to evolve the equations.
We use the 4th order accurate central finite differencing to evaluate spatial derivatives
on the interior grid points, while near the null boundary we use 4th order accurate skewed finite differencing.
We use a uniform grid. 
Constraints are monitored but are not used to correct the evolution other than that which was already used to modify the evolution equations.

The CFL condition puts an upper bound on the time-step:
\be
	 |\Delta T| < \frac{\Delta X}{\text{max speed}},
\ee
where
\be
	\text{max speed} = \max \left\{ \frac{N \ex}{\partial_X r} + \left| \frac{\partial_T r}{\partial_X r} \right| \right\}.
\ee
Note that $\Delta T$ is negative for backward evolution in the second stage.

\section{Examples} \label{sec:ex}

We now consider some example spacetimes and show the accuracy of the code. 
In particular, we focus on cosmological void models, which are spherically symmetric spacetimes with a large-scale under-density at the centre at recent epoch.
They have been considered to explain the dimming of SNIa without introducing a dark-energy component (see, e.g., \cite{Clarkson:2012bg} for a review, and references therein). 

\begin{figure}[t]
 \begin{minipage}[htb]{0.3\textwidth}
   \centering
   \includegraphics[width=\textwidth]{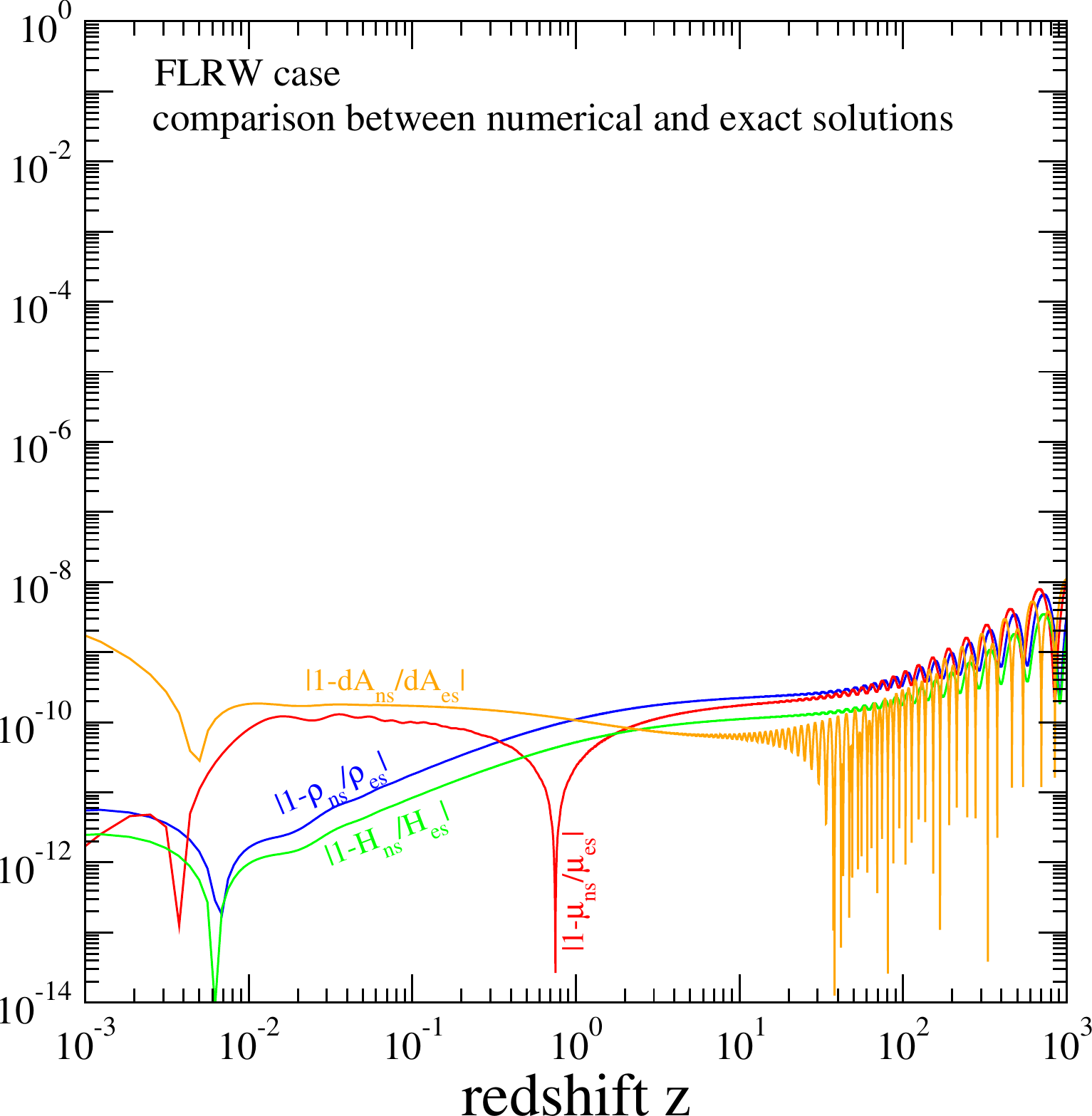}
 \end{minipage}
 \ \hspace{1mm} \
 \begin{minipage}[htb]{0.3\textwidth}
   \centering
   \includegraphics[width=\textwidth]{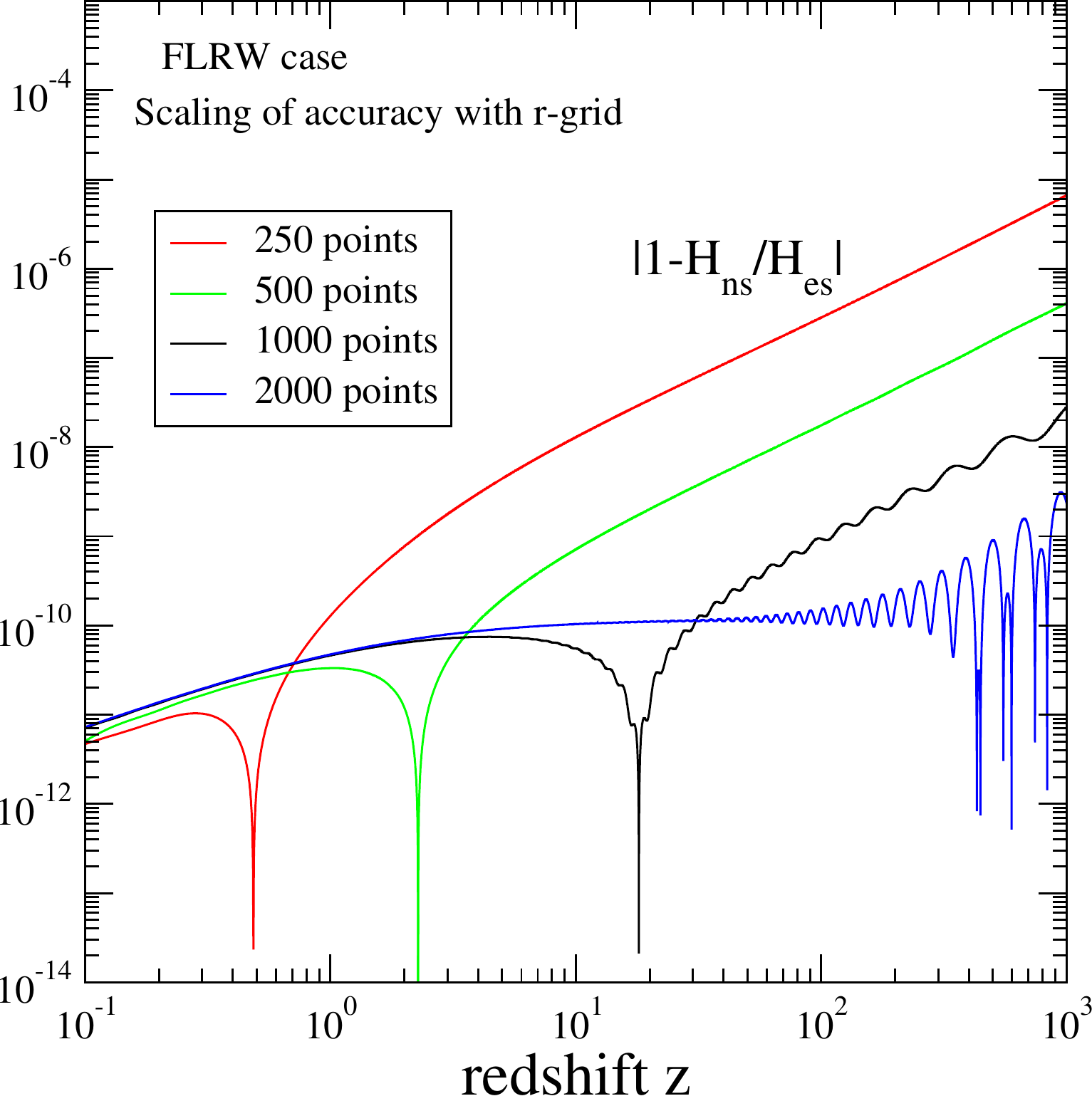}
 \end{minipage}
 \ \hspace{1mm} \
 \begin{minipage}[htb]{0.3\textwidth}
   \centering
   \includegraphics[width=\textwidth]{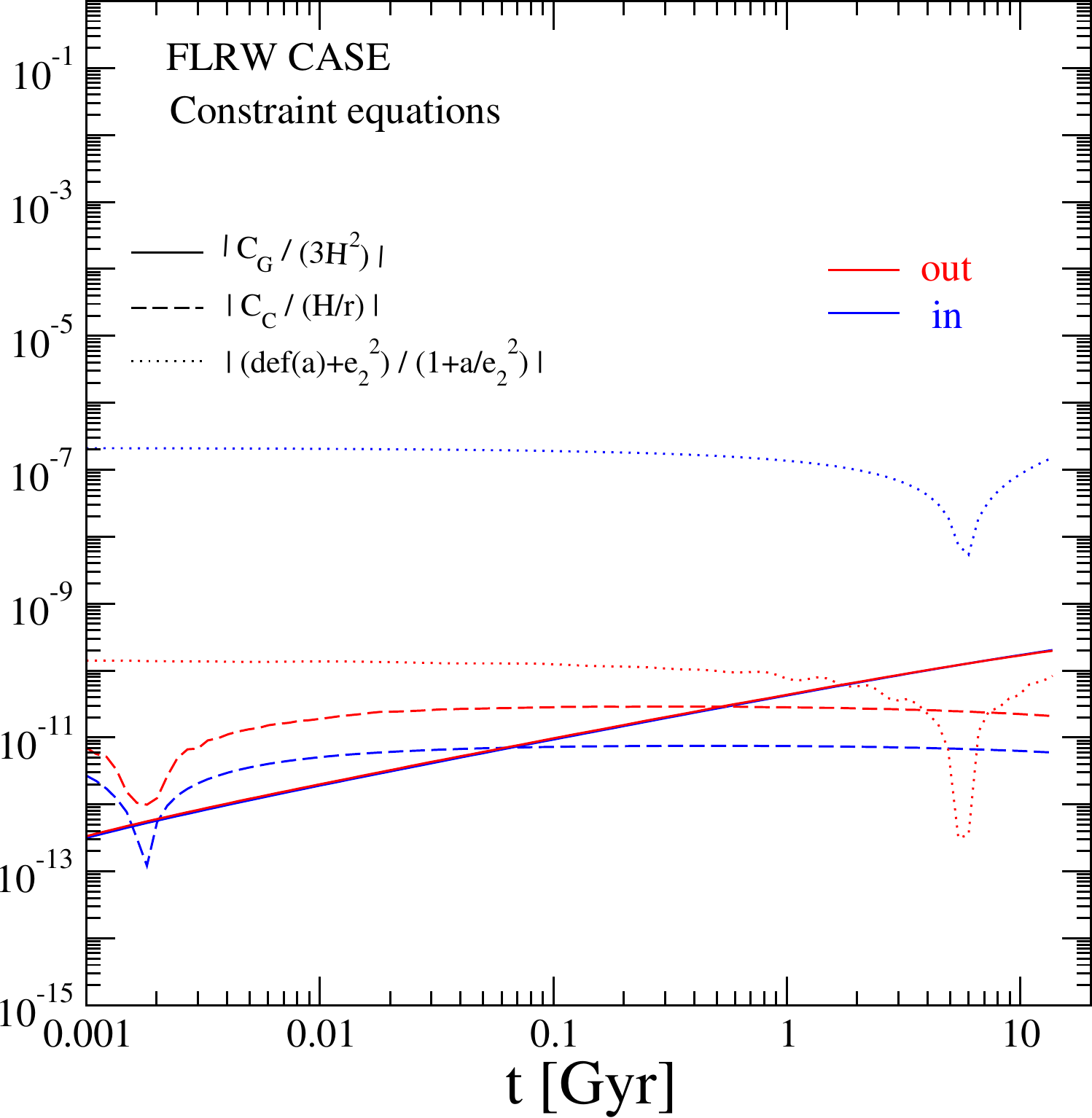}
 \end{minipage}
    \caption{{\bf Homogeneous FLRW limit.} {\it Left:} Comparison between numerical (ns) and analytic exact (es) solutions in the FLRW case. We plot $|\delta_i|=|1-f_{ns}^i/f_{es}^i|$ as a function of redshift, with $f^i$ being matter density $\rho$, radiation density $\mu$, Hubble rate $H$, and area distance $dA$. {\it Left:} Same of left panel for $H(z)$ using different numbers of points for the $r$-grid. {\it Right:} Numerical output for the constraints in Eqs.~(\ref{numcon_start})--(\ref{numcon_end}), which should be zero. They are shown for central and asymptotic regions.}
\label{fig:FLRW}
 \end{figure}

%%%%%%%%%%%%%%%%%%%%%%%%%%%%%%%%%%%%%%
The specification of initial conditions in those models goes as follows. 
Since inhomogeneities are larger than a few hundreds of Mpc in size (in comoving units), while the sound horizon, setting the largest scale
seen in the pre-decoupling part, is around 150 Mpc, the evolution up to a time $t_\text{early}$ (pre-decoupling) is described by a different FLRW model along each worldline. 
Indeed, as long as the gradient of inhomogeneities is small in any causally connected patch of the Universe,
the evolution of horizon-size regions can be described by the mean of the FLRW equations.

At $t_\text{early}$, we set the shear $\sigma_+$ to zero, and specify $\Omega_m(r)$ and $\Omega_r(r)$.
For example, one can choose Gaussian profiles:
\be
         \Omega_m(t_\text{early},r) = (\Omega_m)_\text{out}(t_\text{early})
     + [ (\Omega_m)_\text{in}(t_\text{early}) - (\Omega_m)_\text{out}(t_\text{early}) ]  e^{-r^2/w_m^2}
\label{profile}
\ee
and similarly for $\Omega_r(t_\text{early},r)$, where the label ``in" is used for the worldline at the origin, while ``out" stands for spatial infinity (with $r$ being the radial coordinate and $w$ being the width of the inhomogeneity). 
In this case, initial conditions reduce to specifying 6 parameters at early times: 
$\{ (\Omega_m)_\text{in}, (\Omega_r)_\text{in}, (\Omega_m)_\text{out}, (\Omega_r)_\text{out}, w_m, w_r \}(t_\text{early})$. 
However, we tested the code with other more complicated profiles (see below).

We compute $H=H_{\|}= H_{\perp}$ at $t_\text{early}$ by integrating between $t=0$ and $t=t_\text{early}$ with FLRW evolution along each worldline and (an important condition) setting the bang time to zero.
We fix the gauge taking $\ex$ to be constant. The other functions are then derived by solving the constraint equations.

%%%%%%%%%%%%%%%%%%%%%%%%%%%%%%%%%%%%%%

\subsection{The homogeneous FLRW limit}

In Fig.~\ref{fig:FLRW} we test the accuracy of the code in the limit where both matter and radiation are homogeneous.
In this case, we can make use of the standard analytic solution of Einstein equations (setting the cosmological constant $\Lambda=0$) to compute relevant quantities and compare to the code output.
In Fig.~\ref{fig:FLRW}a, $\rho$, $\mu$, $H$, and $d_A$ are computed along the lightcone of the central observer. The agreement is at a level below $10^{-8}$ up to $z=10^3$ (and a significant part of the errors is due to the redshift interpolation rather than inaccuracy of the solution). In this example we choose the today parameters to be $h_0=0.7$, $\Omega_m=0.7$ with the radiation set by $T_{CMB}=2.725$ K and the effective number of relativistic degrees of freedom given by $N_\text{eff}=3.04$. 

As mentioned above, the solution is 4th order accurate. We explicitly show this in Fig.~\ref{fig:FLRW}b by plotting $H(z)$ obtained considering different numbers of $r$-grid points. The scaling of the inaccuracy has to be constant in time for a stable solution and given by $(N_1/N_2)^p$, where $N_i$ is the number of points in the $r$-grid and $p$ is the order of the accuracy.
It is clear that $p\simeq 4$ with a fairly constant ratio between the different curves. With a few thousand points we reach the current noise level of the code (see blue curve). 
 
The constraint equations (\ref{numcon_start})--(\ref{numcon_end}) are satisfied at a level better than $10^{-6}$ at all times, see Fig.~\ref{fig:FLRW}c. 
We show them along the central worldline (which is the one typically having largest errors due to spurious effects related to the boundary conditions) and along a worldline far from the centre in an ``asymptotic" region (where the 
errors are typically the smallest).

\subsection{Pure-dust LTB versus two-fluid solution}\label{sec:ltb}

\begin{figure}[t]
 \begin{minipage}[htb]{0.3\textwidth}
   \centering
   \includegraphics[width=\textwidth]{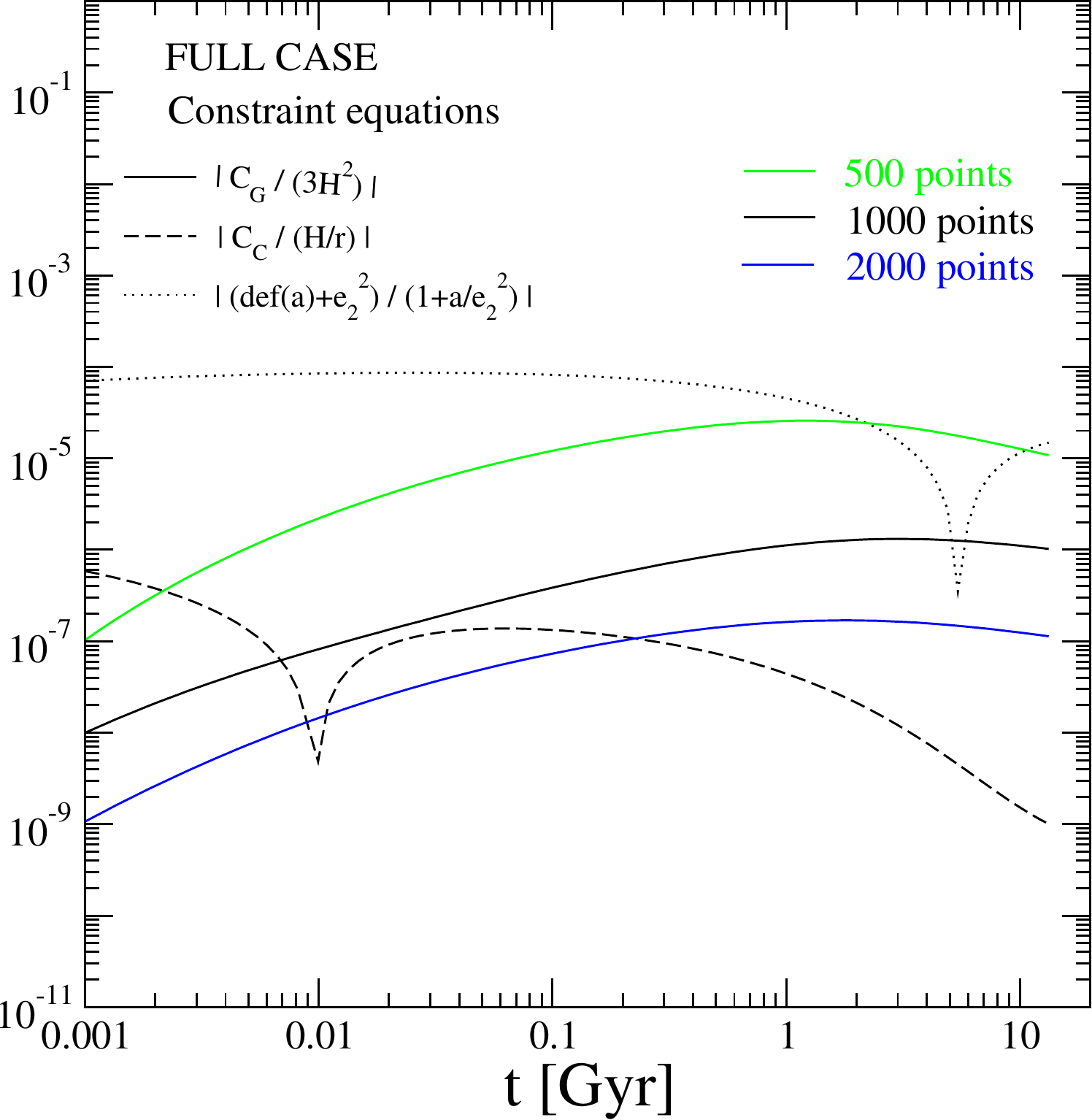}
 \end{minipage}
 \ \hspace{1mm} \
 \begin{minipage}[htb]{0.3\textwidth}
   \centering
   \includegraphics[width=\textwidth]{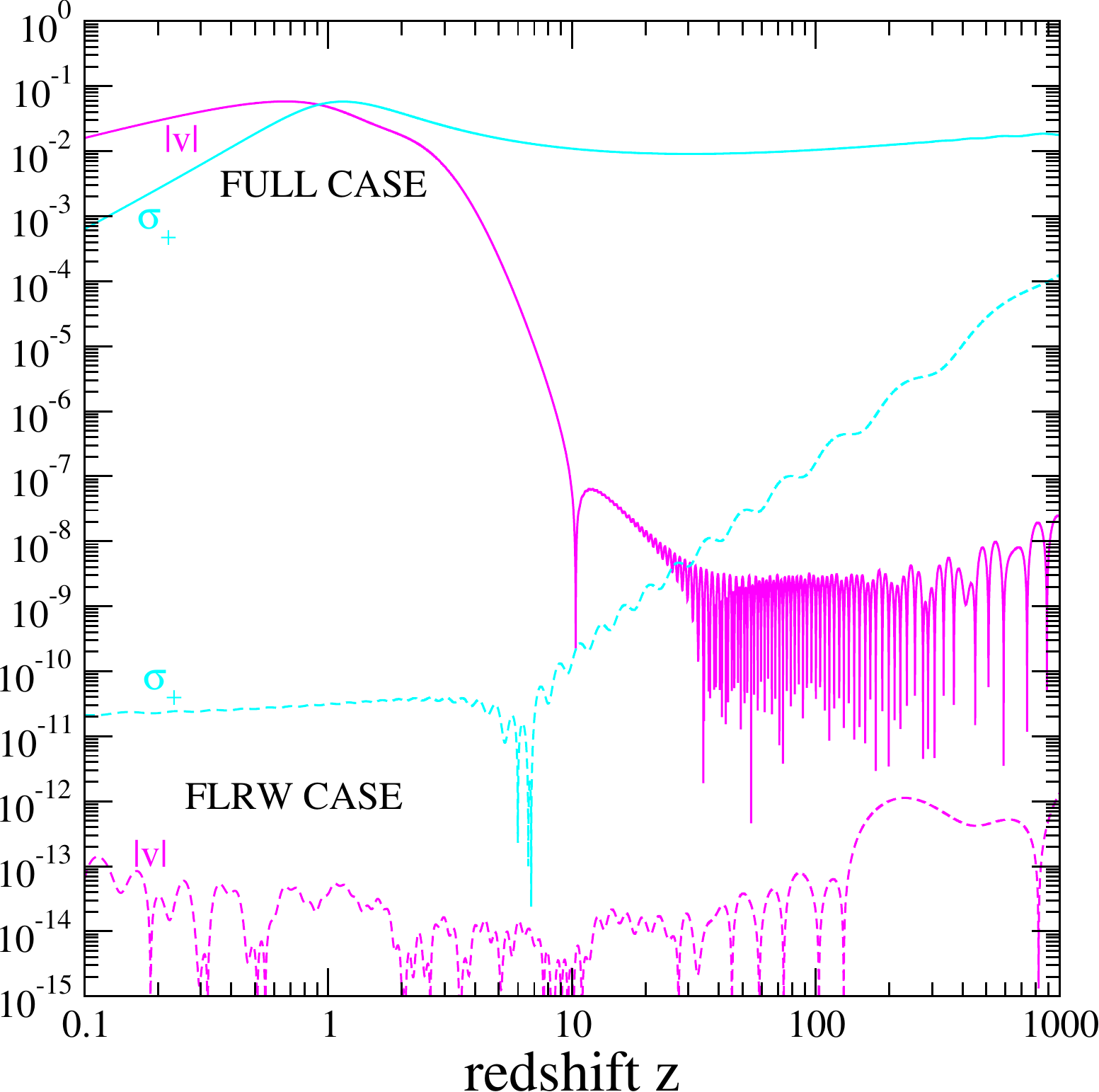}
 \end{minipage}
 \ \hspace{1mm} \
 \begin{minipage}[htb]{0.3\textwidth}
   \centering
   \includegraphics[width=\textwidth]{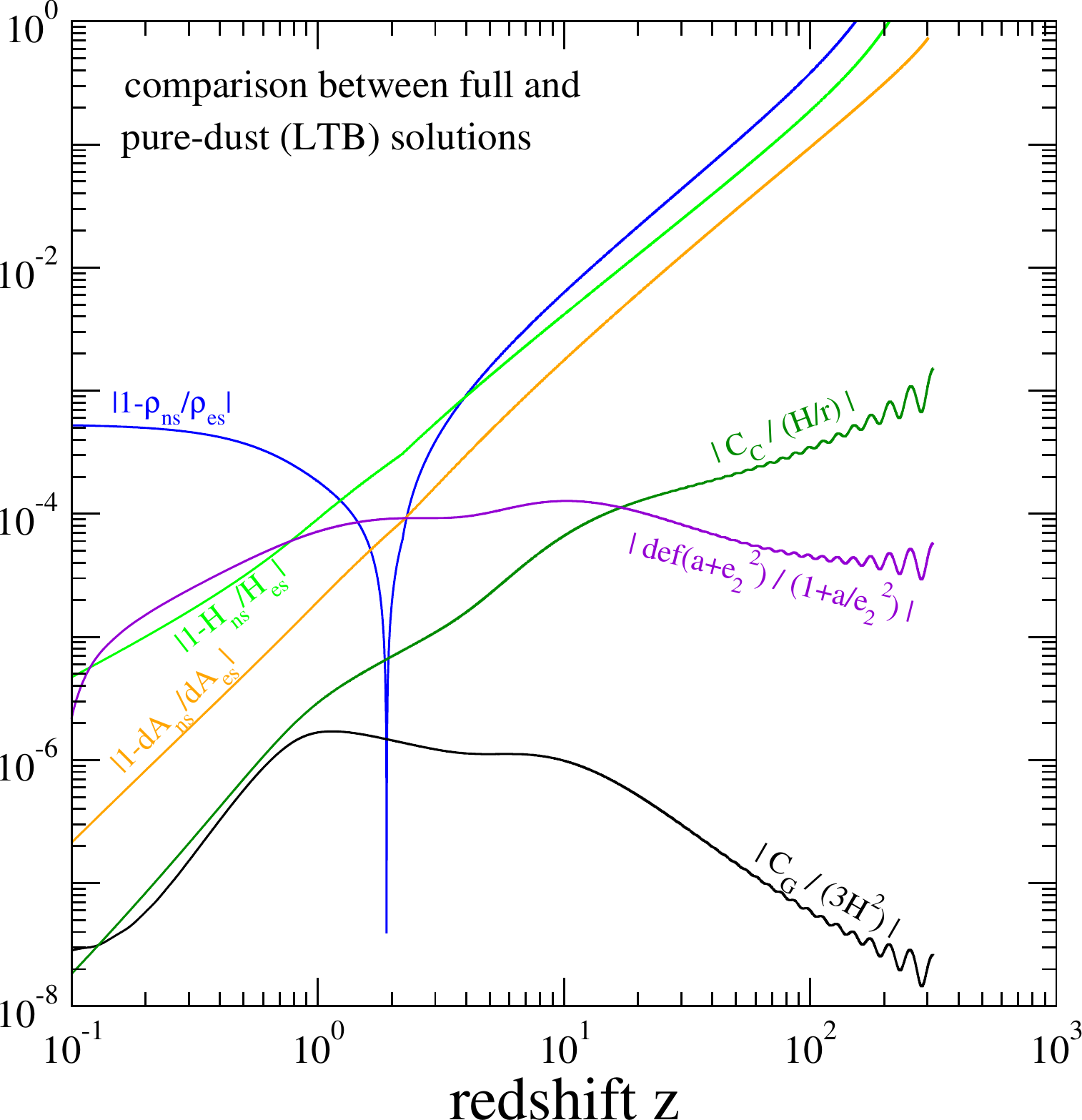}
 \end{minipage}
    \caption{{\bf Pure-dust LTB versus the two-fluid solution.} {\it Left:} Constraints in Eqs.~(\ref{numcon_start})--(\ref{numcon_end}) computed along the central worldline. The scaling of the Gauss constraint with the number of points in the r-grid is shown (similar results can be obtained for the other two constraints). {\it Central:} Shear $\sigma_+$ and peculiar velocity $v$ as a function of redshift for the FLRW and the inhomogeneous (including radiation) models quoted in the text. {\it Right:} Comparison between full (including radiation) and pure dust (LTB) solutions. For the latter, we show also the constraints in Eqs.~(\ref{numcon_start})--(\ref{numcon_end}) which are explicitly violated, although at a level much lower than the actual differences between the two solutions.}
\label{fig:comparison}
 \end{figure}

Now we want to repeat the accuracy analysis in an inhomogeneous scenario, and see how the full solution with two fluids compares to pure-dust LTB models normally adopted in the literature.
For the sake of concreteness we choose a model (other examples give similar results) which leads to parameters on the today surface being $h_0^{(in)}= 0.7$, $\Omega_m^{(out)}\simeq 0.7$, $\Omega_m^{(in)}\simeq 0.16$, $T_0^{(in)}= 2.725$ K, and $T_0^{(out)}\simeq 2.8$ K.
The inhomogeneity profile is modelled through a Gaussian with a width corresponding to $z=1$. 

The constraint equations (\ref{numcon_start})--(\ref{numcon_end}) are satisfied at a level better than $10^{-4}$, as shown for the central worldline in Fig.~\ref{fig:comparison}a.
The accuracy is clearly lower than in the homogeneous case, but the scaling is again constant and not far from 4th order (see ratio between solid lines).

Physical quantities which become non-zero in the inhomogeneous scenario are the shear $\sigma_+$ and the peculiar velocity $v$. They indeed get large values in this model, peaking at redshift corresponding to the size of the inhomogeneity.
In the homogeneous scenario, on the other hand, the shear has to be zero and radiation is comoving with matter (i.e., $v=0$). See Fig.~\ref{fig:comparison}b.
The numerical error on $\sigma_+$ and $v$ never becomes relevant, although the growing of $\sigma_+$ suggests that some improvements in the computation of $z$-scalings can be made (indeed the right part of the plot is where we set initial conditions, thus the inaccuracy should be the smallest).

In Fig.~\ref{fig:comparison}c we compare such two-fluid solution to a pure dust solution.
As we mentioned above, in order to make equations more tractable, we specify initial conditions such that $\ex$ was set to a constant. It implies we are using hyperspherical coordinates.
On the other hand this makes it inconvenient to compare with LTB models, whose common choice of reduced-circumference polar coordinates means $\ex$ depends on $r$ (e.g., $\ex = \sqrt{1-kr^2}$ for FLRW).
Therefore, it is not straightforward to simplify this model to an LTB model.
Moreover, simply setting $\mu$ and $v$ to zero would violate the Codazzi constraint.
In order to satisfy the Codazzi constraint, we should set $H$ to constant, which means
we have no control over the bang time.

We proceed with specifying for the pure dust case. 
We use the same input as for the two-fluid model except for replacing $\rho_\text{LTB}=\rho_\text{full}+\mu_\text{full}$ in the initial condition at $z=0$ and setting $\mu=v=\dot v=0$ in all equations. 
In this way we explicitly violate constraints in Eqs.~(\ref{numcon_start})--(\ref{numcon_end}). However, this violation is much smaller than the actual differences between the two solutions as shown. 
We plot $|\delta_i|=|1-f_\text{full}^i/f_\text{LTB}^i|$ as a function of redshift, with $f^i$ being matter density $\rho$, Hubble rate $H$, and area distance $d_A$. The LTB solution is a good approximation up to $z\lesssim10$, 
while at $z=100$ we find $\delta_i\gtrsim10\%$. This is larger than the radiation contribution to the energy density at such redshift (which is about 3\%). Those results are similar to what found in Fig. 12 of 
\cite{Clarkson:2010ej}, telling us that the computation of the full solution by matching an LTB spacetime to an FLRW spacetime at $z\sim100$ (as often done in the literature) might not be accurate enough to be exploited for CMB 
computations (this topic will be  discussed in a forthcoming paper).

\subsection{Void models with matter and radiation inhomogeneities}

\begin{figure}[t]
 \begin{minipage}[htb]{0.45\textwidth}
   \centering
   \includegraphics[width=\textwidth]{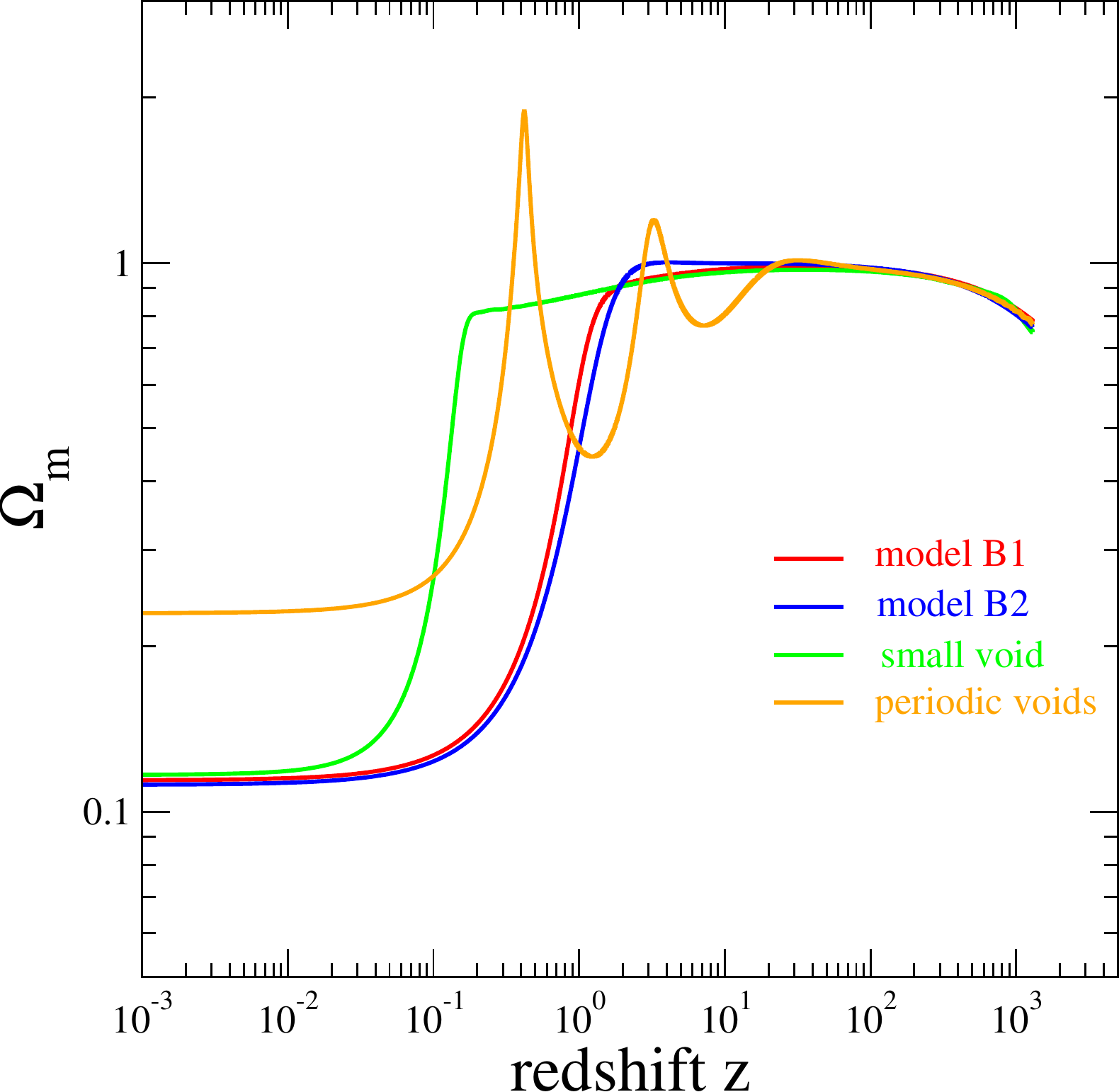}
 \end{minipage}
 \ \hspace{1mm} \
 \begin{minipage}[htb]{0.45\textwidth}
   \centering
   \includegraphics[width=\textwidth]{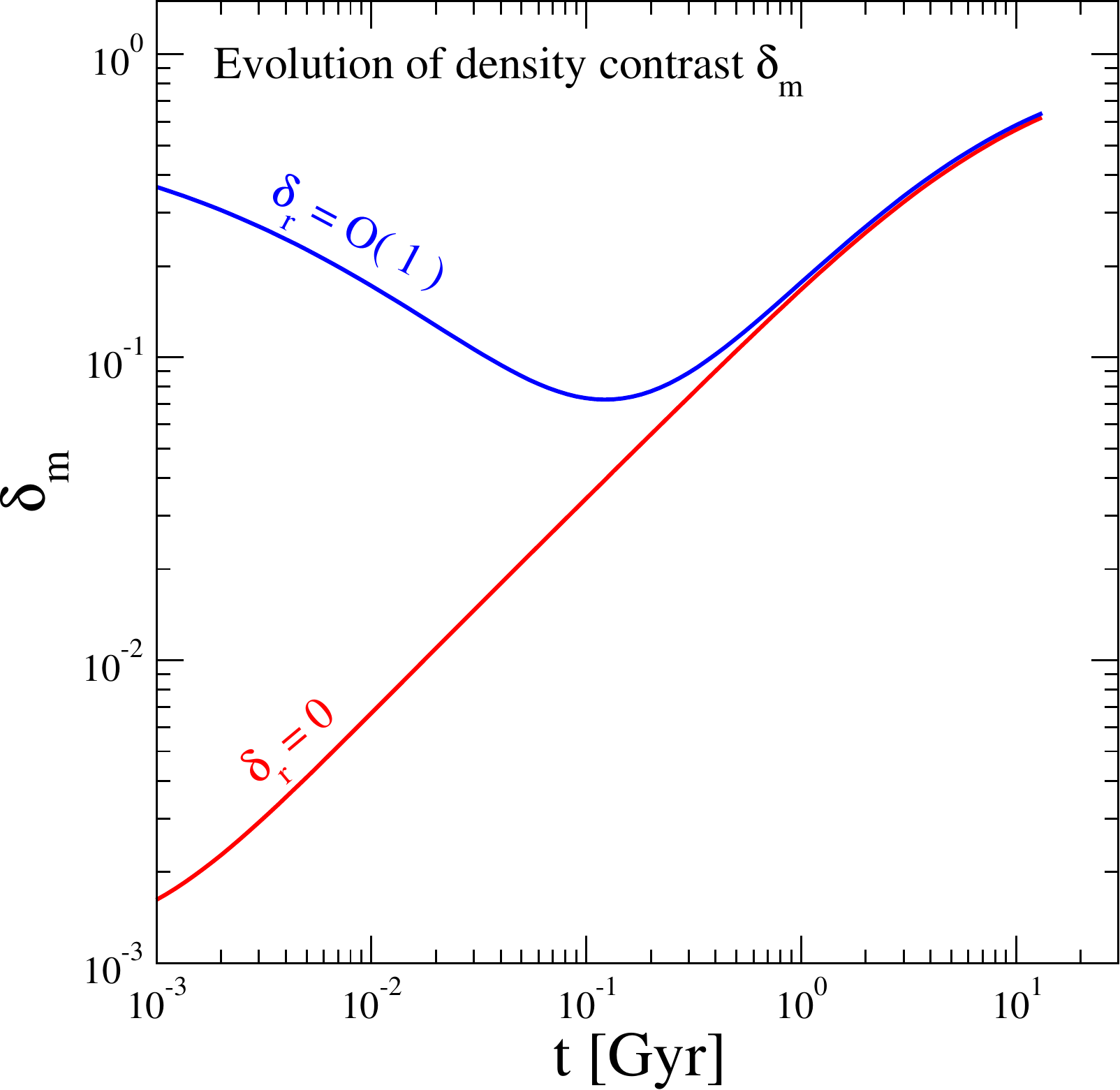}
 \end{minipage}
    \caption{{\bf Void models with matter and radiation inhomogeneities.} {\it Left:} Matter density along the past-light cone of the central observer for different initial profiles. Gaussian profiles with Gpc inhomogeneity only in matter and in both matter and radiation are shown in red and blue, respectively. A small Gaussian void is shown in green, while a profile with a sinusoidal initial density profile is in orange.  {\it Right:} Comparison of the evolution of the matter density contrast between the central and the asymptotic regions for the Gpc-voids of left panel (Gaussian profiles).
The radiation is taken to be homogeneous (red) and with an $\mathcal{O}(1)$ inhomogeneity (blue), at early times. }
\label{fig:inhomrad}
 \end{figure}

\begin{figure}[t]
 \begin{minipage}[htb]{0.45\textwidth}
   \centering
   \includegraphics[width=\textwidth]{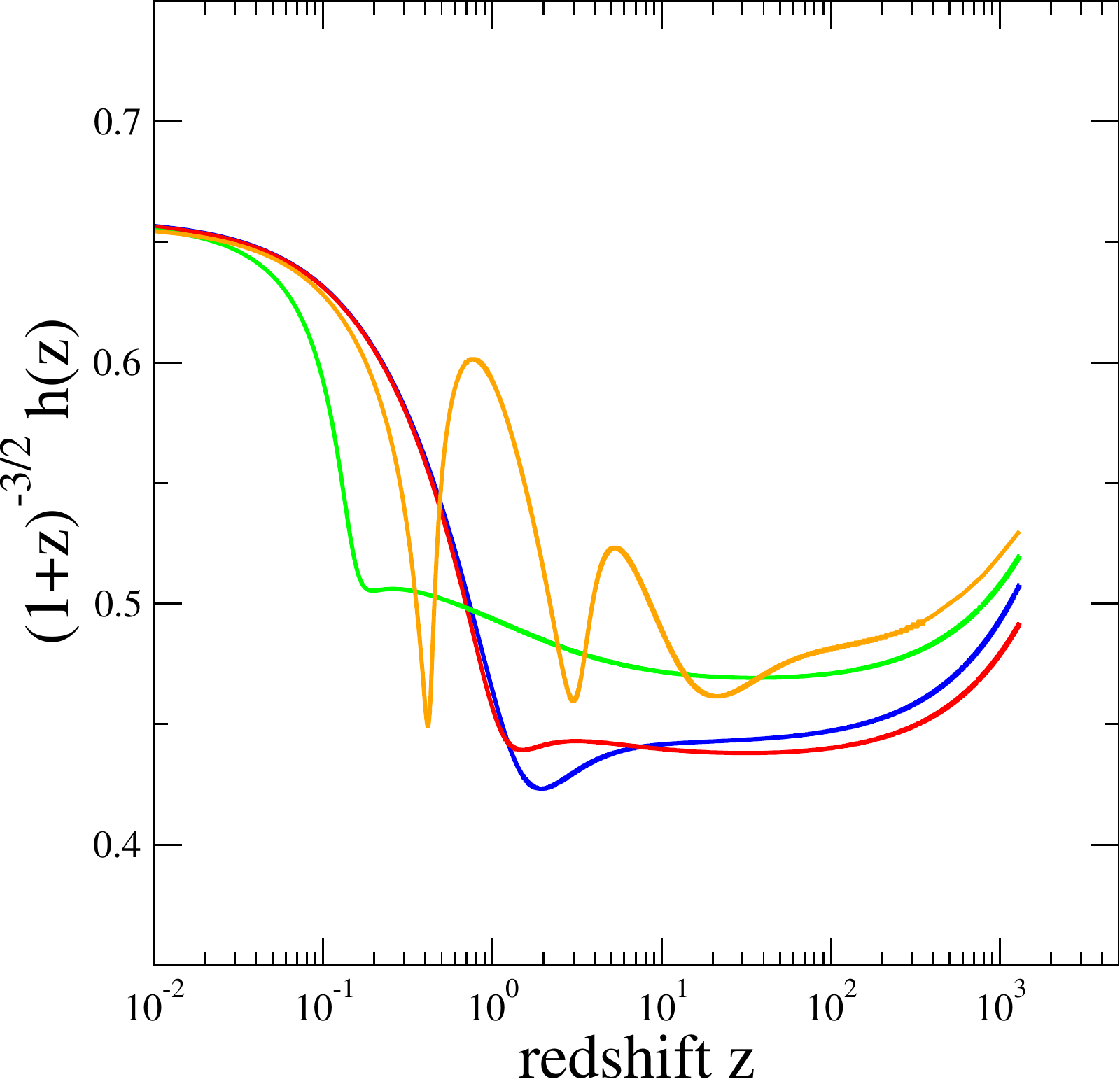}
 \end{minipage}
 \ \hspace{1mm} \
 \begin{minipage}[htb]{0.45\textwidth}
   \centering
   \includegraphics[width=\textwidth]{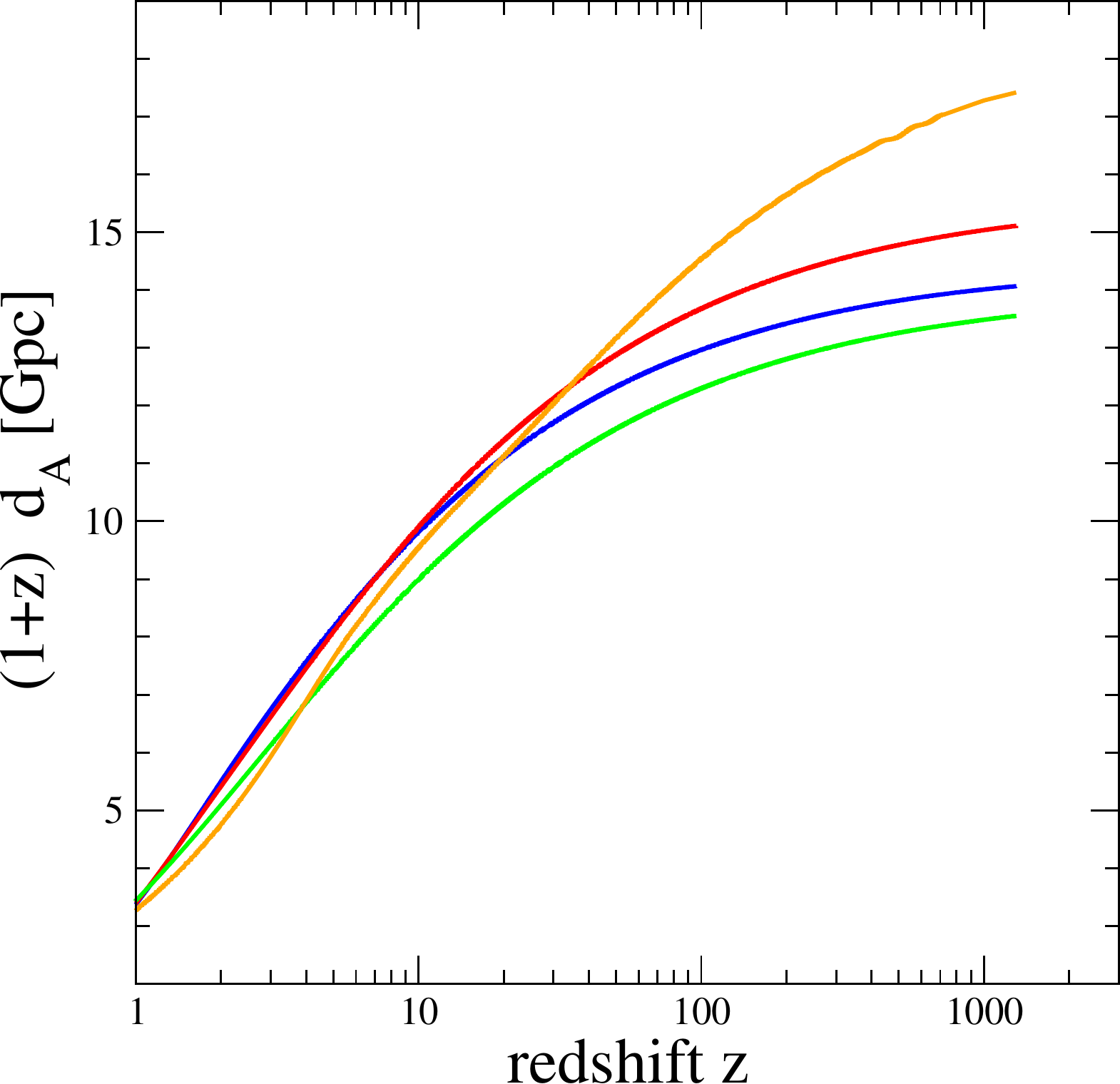}
 \end{minipage}
    \caption{{\it Left:}  Hubble rate along past-light cone for the same models as in the left panel of Fig.~\ref{fig:inhomrad}.  {\it Right:} Same as in the left panel, but for the angular diameter distance. Note the difference in the comoving distance of $\sim1$\,Gpc at $z\sim1000$ brought on by changing the radiation profile in the initial conditions. }
\label{fig:modelsvsz}
 \end{figure}

In Figs.~\ref{fig:inhomrad} and \ref{fig:modelsvsz} we explore the evolution of the matter density, Hubble rate, and angular diameter distance for a few examples of the full solution.
We set the radiation and matter spatial profiles at initial time $t^{(\text{in})}=10^5$ s.
The simplest case is a Gaussian as in Eq.~(\ref{profile}).
We consider two models having a few Gpc width and provide identical parameters at $z=0$ (listed in Section~\ref{sec:ltb}), the first with homogeneous radiation at early times (red), i.e. with 
$\Omega_r^{(\text{in})}/\Omega_r^{(\text{out})}=1$, and the second with $\Omega_r^{(\text{in})}/\Omega_r^{(\text{out})}=2$ (blue). The first case sketches the class of models often considered in the literature to fit SNIa data without dark-energy, while a void model with early-times inhomogeneity in the radiation profile can potentially fit also BBN~\cite{Regis:2012iq} and CMB~\cite{Clarkson:2010ej} data.

It's interesting to note in Fig.~\ref{fig:inhomrad}b how in the first case (adiabatic scenario) the matter density contrast $\delta_m=|1-\rho_m^{(\text{in})}/\rho_m^{(\text{out})}|$ is highly homogeneous at early times, while in the second case the picture starts inhomogeneous (isocurvature scenario). 
They are set to give the same physical picture today and have the same redshift evolution for the matter profile, but the $H$ and $d_A$ at early times can be slightly different (see Fig.~\ref{fig:modelsvsz}) because of the radiation inhomogeneity. 

Alongside, we consider a case where the width is one order of magnitude smaller (i.e. about 300 Mpc) and a sinusoidal function with periodic over/under-densities $\Omega_i(t_\text{early},r) = (\Omega_i)_\text{out}(t_\text{early})
     + [ (\Omega_i)_\text{in}(t_\text{early}) - (\Omega_i)_\text{out}(t_\text{early}) ]\,\cos(r/w)^2$ with $i=m,\,r$. Note that in the latter case, since obviously the matter density parameter $\Omega_m$ tends to 1 in the matter dominated epoch, the oscillations look damped as a function of $z$.

\section{Conclusion} \label{sec:conc}

We have developed a numerical scheme for simulating spherically symmetric spacetimes with a comoving dust and a tilted radiation fluid. 
This scheme avoids boundary conditions by using a new zooming technique improved from that of~\cite{art:Limetal2009} with null boundaries. 
The numerical evolution equations are given by (\ref{numevo_start})--(\ref{numevo_end}) and (\ref{dtr}).
Equation (\ref{dtr_first}) is used for the first stage, and (\ref{dtr_second}) for the second.
The second stage contains one additional evolution equation (\ref{dt_epsilon}) for the redshift.
Using fourth-order Runge-Kutta method and fourth-order finite differencing to evaluate spatial derivatives, the scheme is fourth-order convergent.
Stability of the scheme is not proven, but numerical errors can be monitored by plotting the constraints (\ref{numcon_start})--(\ref{numcon_end}).
Some examples were given to demonstrate the numerical scheme, from which we have confirmed stability.
We then considered models with radiation and compared them to the pure dust LTB case, which demonstrates the importance of including radiation for the evaluation of the area distance at large redshift. We then showed how radiation inhomogeneity can also affect the model, and quantities such as the area distance at large redshift. 

The key limitation of the model is that we have approximated the radiation as a perfect fluid, effectively truncating the Boltzmann hierarchy at the monopole term. When the shear is significant this induces corrections to the 
radiation density $\mathcal{O}(\mu\sigma^2)$ (see, e.g. Eqs. (86) and (89) in~\cite{Maartens:1998xg}). The effects associated with this are interesting to investigate in their own right, and are also important for accurately 
modelling a spherical object with matter and radiation. Further work is also needed to treat the baryons accurately by adding in the coupling to the photons through Compton scattering. 
In a future development, we will examine these effects on the constraints on void models, and in particular the combined bounds from the CMB+$H_0$ which significantly constrain pure dust models.

\section*{Acknowledgements}
We would like to thank A. Nagar,  S. Perez Bergliaffa and F. Teppa Pannia for useful discussions.
MR acknowledges the Research Grant funded by the Istituto Nazionale di Fisica Nucleare within the {\sl Astroparticle Physics Project} (INFN grant code: FA51).

%\bibliography{cites}

\begin{thebibliography}{99}

%\cite{Bolejko:2010zz}
\bibitem{Bolejko:2010zz} 
Bolejko, K., 
Krasi{\'n}ski, A., Hellaby, C., 
\& C{\'e}l{\'e}rier, M.-N., Structures in the Universe by Exact Methods: Formation, Evolution, Interactions.~ Cambridge Monographs on Mathematical Physics, Cambridge University Press (2009)

%\cite{Marra:2011zp}
\bibitem{Marra:2011zp}
  V.~Marra and M.~Paakkonen,
  %``Exact spherically-symmetric inhomogeneous model with n perfect fluids,''
  JCAP {\bf 1201} (2012) 025
  [arXiv:1105.6099 [gr-qc]].
  
%\cite{Marra:2012pj}
\bibitem{Marra:2012pj} 
  V.~Marra, M.~Paakkonen and W.~Valkenburg,
  %``Uncertainty on w from large-scale structure,''
  Mon.\ Not.\ Roy.\ Astron.\ Soc.\  {\bf 431}, 1891 (2013)
  [arXiv:1203.2180 [astro-ph.CO]].  
 
%\cite{Pitrou:2008ak}
\bibitem{Pitrou:2008ak} 
  C.~Pitrou, J.~-P.~Uzan and F.~Bernardeau,
  %``Cosmic microwave background bispectrum on small angular scales,''
  Phys.\ Rev.\ D {\bf 78}, 063526 (2008)
  [arXiv:0807.0341 [astro-ph]].

%\cite{Clarkson:2012bg}
\bibitem{Clarkson:2012bg} 
  C.~Clarkson,
  %``Establishing homogeneity of the universe in the shadow of dark energy,''
  Comptes Rendus Physique {\bf 13}, 682 (2012)
  [arXiv:1204.5505 [astro-ph.CO]].

%\cite{Moss:2010jx}
\bibitem{Moss:2010jx} 
  A.~Moss, J.~P.~Zibin and D.~Scott,
  %``Precision Cosmology Defeats Void Models for Acceleration,''
  Phys.\ Rev.\ D {\bf 83}, 103515 (2011)
  [arXiv:1007.3725 [astro-ph.CO]].
  %%CITATION = ARXIV:1007.3725;%%
  %50 citations counted in INSPIRE as of 01 Jul 2013

%\cite{Clarkson:2010ej}
\bibitem{Clarkson:2010ej} 
  C.~Clarkson and M.~Regis,
  %``The Cosmic Microwave Background in an Inhomogeneous Universe - why void models of dark energy are only weakly constrained by the CMB,''
  JCAP {\bf 1102}, 013 (2011)
  [arXiv:1007.3443 [astro-ph.CO]].

%\cite{Lara:2006cd}
\bibitem{Lara:2006cd}
  J.~F.~Lara, T.~Kajino and G.~J.~Mathews,
  %``Inhomogeneous big bang nucleosynthesis revisited,''
  Phys.\ Rev.\ D {\bf 73} (2006) 083501
  [astro-ph/0603817].

%\cite{vanElst:1996dr}
\bibitem{vanElst:1996dr} 
  H.~van Elst and C.~Uggla,
  %``General relativistic (1+3) orthonormal frame approach revisited,''
  Class.\ Quant.\ Grav.\  {\bf 14}, 2673 (1997)
  [gr-qc/9603026].

\bibitem{book:WainwrightEllis1997}
J. Wainwright and G.~F.~R. Ellis, {\em Dynamical systems in cosmology}
  ({C}ambridge {U}niversity {P}ress, Cambridge, 1997).

%\cite{Coley:2008zz}
\bibitem{Coley:2008zz} 
  A.~A.~Coley, S.~Hervik and W.~C.~Lim,
  %``General relativistic tilt and dark energy,''
  AIP Conf.\ Proc.\  {\bf 1083}, 65 (2008).

%\cite{Coley:2008qd}
\bibitem{Coley:2008qd} 
  A.~A.~Coley, W.~C.~Lim and G.~Leon,
  %``Spherically symmetric cosmology: Resource paper,''
  arXiv:0803.0905 [gr-qc].

\bibitem{book:Stephanietal2003}
H. Stephani {\it et~al.}, {\em Exact solutions of {E}instein's field equations,
  second edition} ({C}ambridge {U}niversity {P}ress, Cambridge, 2003).

%\cite{Alcubierre:2004gn}
\bibitem{Alcubierre:2004gn} 
  M.~Alcubierre and J.~A.~Gonzalez,
  %``Regularization of spherically symmetric evolution codes in numerical relativity,''
  Comput.\ Phys.\ Commun.\  {\bf 167}, 76 (2005)
  [gr-qc/0401113].
  
\bibitem{art:Limetal2009}
W.~C. Lim, L. Andersson, D. Garfinkle, and F. Pretorius, Phys. Rev. D {\bf 79},
   103526  (2009).  
  
%\cite{Regis:2012iq}
\bibitem{Regis:2012iq}
  M.~Regis and C.~Clarkson,
  %``Do primordial Lithium abundances imply there's no Dark Energy?,''
  Gen.\ Rel.\ Grav.\  {\bf 44} (2012) 567
  [arXiv:1003.1043 [astro-ph.CO]].

  
  
%\cite{Maartens:1998xg}
\bibitem{Maartens:1998xg} 
  R.~Maartens, T.~Gebbie and G.~F.~R.~Ellis,
  %``Covariant cosmic microwave background anisotropies. 2. Nonlinear dynamics,''
  Phys.\ Rev.\ D {\bf 59}, 083506 (1999)
  [astro-ph/9808163].  




\end{thebibliography}

\end{document}